\documentclass[CEJP,DVI]{cej} % use DVI command to enable LaTeX driver
\usepackage{layout}
\usepackage{amsmath}
\usepackage{textcomp}
\usepackage{hyperref}
\usepackage{epsfig}  %added by KG
\usepackage{wrapfig} %added by	KG

\title{Status and plans of the ion program of NA61 at the CERN SPS}

\articletype{Editorial}

\author{Katarzyna 
Grebieszkow\inst{1}\email{kperl@if.pw.edu.pl} 
(for the NA61/SHINE Collaboration)}

\institute{
     \inst{1} Faculty of Physics, Warsaw University of Technology,\\
        Koszykowa 75, 00-662~Warsaw, Poland
          }

\abstract{

The NA61/SHINE at the CERN SPS is a new experiment to study hadron production in $p+p$, $p+A$, $h+A$ and $A+A$ interactions. The main goal of the NA61 ion program is to explore the phase diagram ($T - \mu_B$) of strongly interacting matter. In particular, we plan to study the properties of the onset of deconfinement and to search for the signatures of the critical point. A two-dimensional scan of the phase diagram will be performed by varying the energy (13$A$-158$A$ GeV) and system size ($p+p$, $Be+Be$, $Ar+Ca$, $Xe+La$) of collisions. This paper summarizes the status and plans of the NA61/SHINE ion program. In particular the detector upgrades, data taking schedule and the first results on spectra and correlations are discussed.}

\keywords{NA61/SHINE experiment \*\ SPS experiment \*\ ion program \*\ critical point \*\ onset of deconfinement}
\pacs{25.75.Ag, 25.75.Gz, 25.75.Dw}

\begin{document}
\maketitle

%% ###################################################################

\section{NA61/SHINE physics program}

The energies of the CERN SPS accelerator cover a very important region of the phase diagram of strongly interacting matter. First, the NA49 experiment \cite{na49_nim} showed that the energy threshold for deconfinement (minimum energy to create a partonic system) is located at low SPS energies ($\sqrt{s_{NN}} \approx 7.6$~GeV or 30$A$ GeV) \cite{na49_kpi}. Second, theoretical calculations suggest that the critical point (CP) is located at energies accessible at the CERN SPS, i.e. $T^{CP} = 162 \pm 2$ MeV, $\mu_B^{CP} = 360 \pm 40 $ MeV) \cite  {fodor_latt_2004} or $(T^{CP}, \mu_B^{CP}) = (0.927(5)T_c, 2.60(8)T_c) = (\sim 157, \sim 441)$ MeV~\cite{lat_2011}.

\begin{figure}
\includegraphics[width=0.5\textwidth]{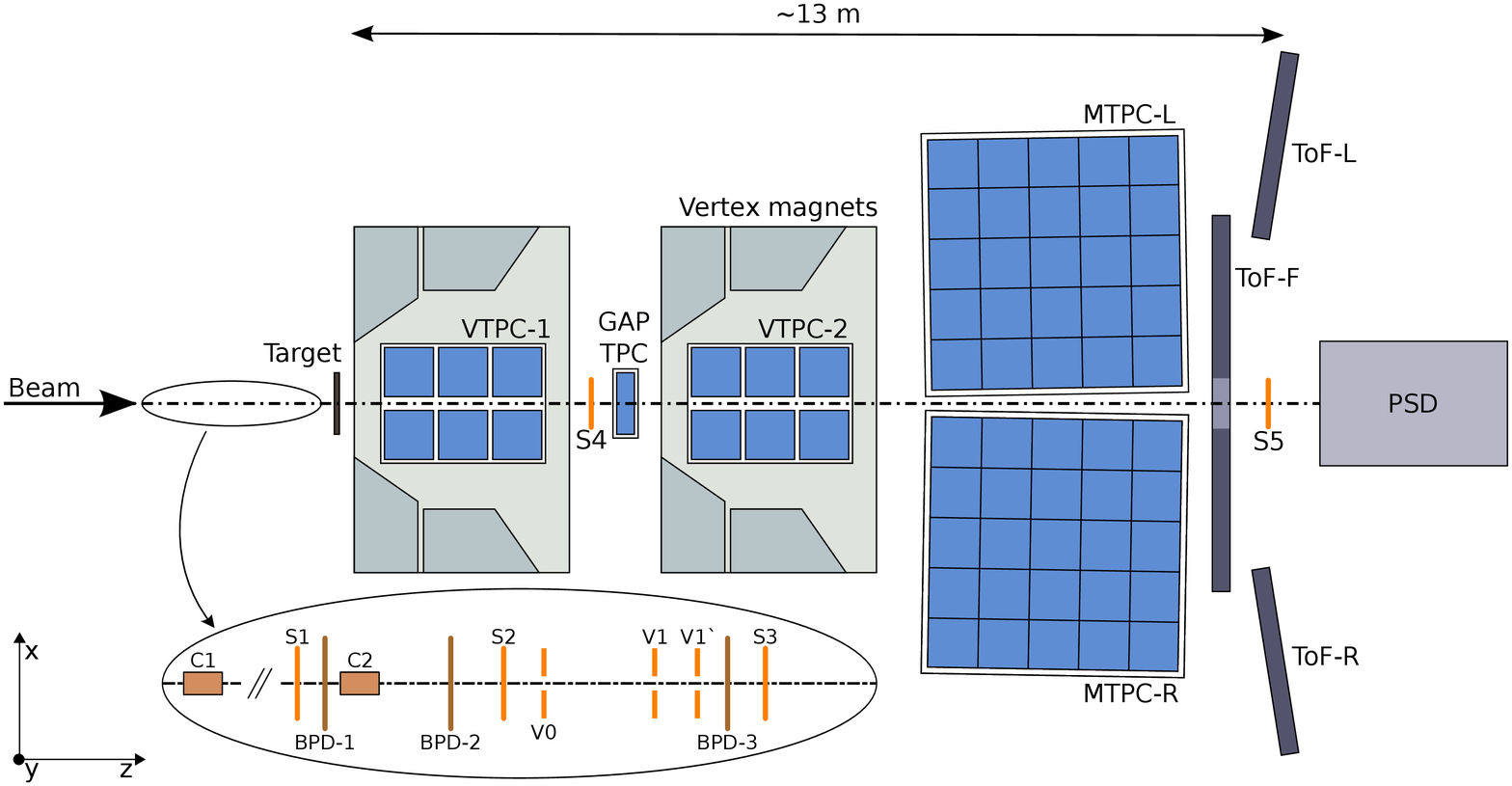}
\includegraphics[width=0.4\textwidth]{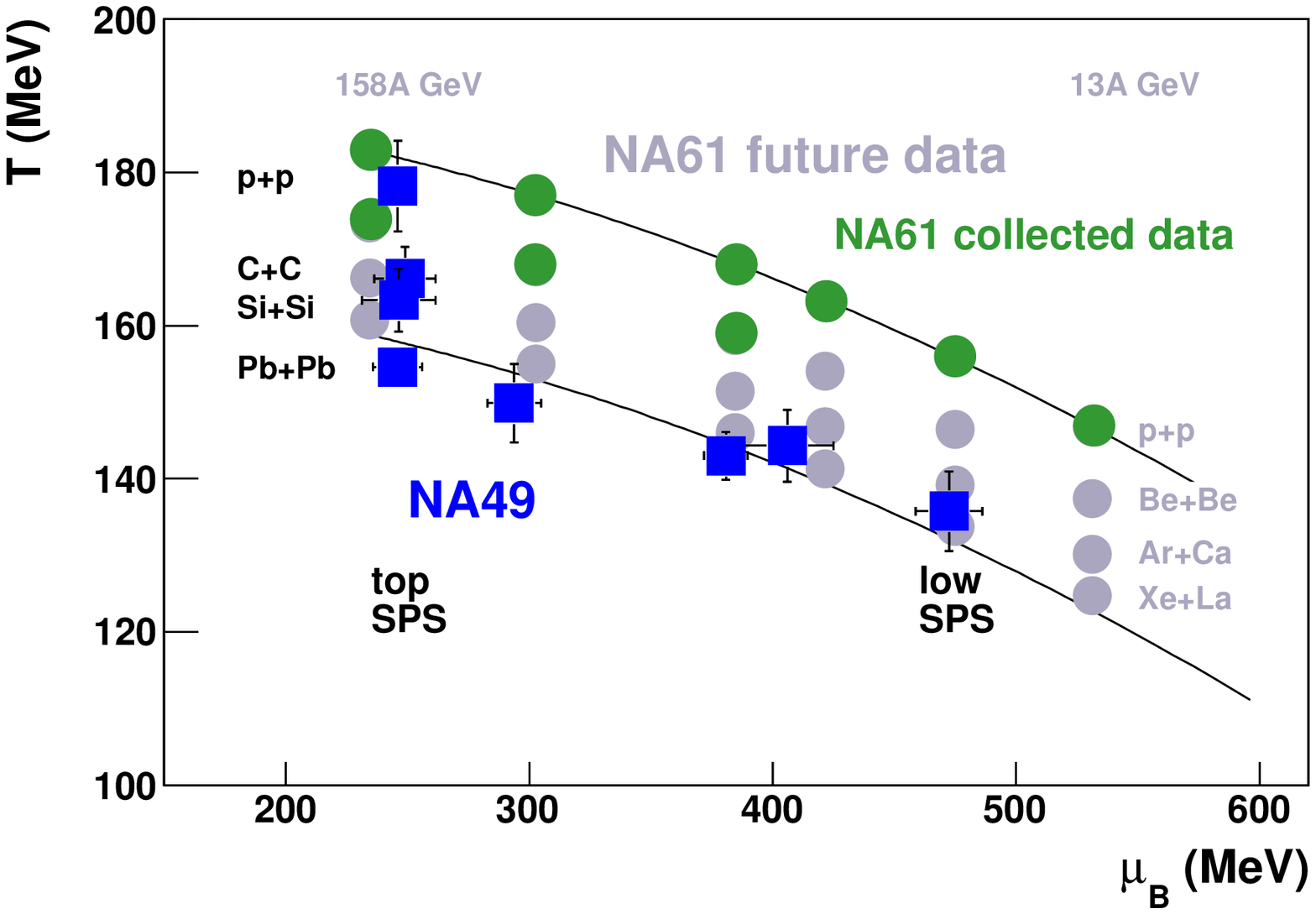}
\vspace{-0.3cm}
\caption{NA61 setup (left). Estimated (NA49) and expected (NA61) chemical 
freeze-out points accordingly to \cite{beccatini} (right). NA49 data indicate \cite{beccatini} that at the top SPS energy $\mu_B$ does not depend on the system size ($C+C$, $Si+Si$, $Pb+Pb$). Therefore the $\mu_B$ values for $p+p$ are also displayed and assumed to be the same as for $Pb+Pb$.}
\label{ex_phase_diagram}
\end{figure}

NA61/SHINE \footnote{SHINE - SPS Heavy Ion and Neutrino Experiment} is a fixed target experiment (see Fig. \ref{ex_phase_diagram} (left)) in the north area of the CERN SPS. It uses the upgraded NA49 \cite{na49_nim} detector, started data taking in 2007, and will study hadron production in $p+p$, $p+A$, $h+A$, and $A+A$ collisions at various energies. A broad experimental program is planned: search for the critical point of strongly interacting matter, study of the properties of the onset of deconfinement (OD), study of high $p_T$ particle production (energy dependence of the nuclear modification factor), and analysis of hadron spectra for the T2K neutrino experiment and for the Pierre-Auger and KASCADE cosmic-ray experiments. In fact, first NA61 results on pion spectra in $p+C$ interactions at 31 GeV are published \cite{pion_paper}. They were already used to improve neutrino beam flux predictions in T2K.
% (it allowed to measure $\vartheta _{13}$ parameter of the neutrino oscilation matrix \cite{theta_13}). 

Within the ion program of NA61 we plan to analyze interactions with proton and ion ($Be$, $Ar$, $Xe$) beams at 13$A$, 20$A$, 30$A$, 40$A$, 80$A$, and 158$A$ GeV. This will allow to cover a wide region of the phase diagram (the expected chemical freeze-out points are shown in Fig. \ref{ex_phase_diagram} (right)). The data on $p+p$ interactions at 13-158 GeV were already recorded in 2009-2011. In the year 2011 we recorded also $Be+Be$ collisions at the three highest energies, the remaining three lower energies will be taken in 2012. In 2012/2014 the energy scan (13$A$ - 158$A$ GeV) of $p+Pb$ collisions is planned. $Ar+Ca$ interactions will be recorded in 2014, and finally the energy scan of $Xe+La$ collisions in 2015 will complete data taking for the NA61/SHINE ion program. Several (2-6) million of events were (or will be)
recorded for each system and each energy.

NA49 showed indications of the onset of deconfinement \cite{na49_kpi} in the energy dependence of the pion yield per wounded nucleon $N_W$ (kink), of the $\langle K^{+} \rangle / \langle \pi^{+} \rangle $ ratio (horn), and of the mean transverse mass or inverse slope parameters of $m_T$ spectra (step) in central $Pb+Pb$ collisions at lower SPS energies (30$A$ GeV). The horn and step structures are now confirmed by the RHIC beam energy scan program (see \cite{kg_sqm2011} for details). The NA61 ion program will study the properties of the OD by looking for the onset of the horn, kink, step in collisions of light nuclei (the structures observed for $Pb+Pb$/$Au+Au$ should vanish with decreasing system size). The NA61 data will also allow to search for the critical point. An increase of the CP signal ({\it hill of fluctuations}) is expected for systems freezing-out near the CP \cite{SRS}. 
%\footnote{CP should be searched above the energy of the onset of deconfinement $E_{CP}>E_{OD} \simeq 30A$ GeV.}. 
Therefore non-monotonic dependence of the CP signal (multiplicity and average $p_T$ fluctuations, intermittency, etc.) on control parameters (energy and ion size) can help to locate the CP. In fact, NA49 already noticed that for central $A+A$ collisions fluctuations of average $p_T$, multiplicity of charged particles and of low mass $\pi^{+}\pi^{-}$ pairs tend to a maximum in Si+Si collisions at 158$A$ GeV \cite{kg_qm09, kg_sqm2011, fotis}. This result is a strong motivation for the NA61/SHINE ion program.

\section{Detector status and main upgrades}

The main devices of the NA61 detector, namely four large volume Time Projection Chambers (TPC)  (two of them - Vertex TPCs (VTPC) - inside superconducting magnets, two others - Main TPCs - downstream of the magnets - symmetric to the beam line) and two Time-of-Flight (ToF) walls were inherited from NA49 \cite{na49_nim}. Several upgrades were completed. In 2007 an additional forward ToF wall was constructed to extend ToF acceptance for particles with $p<3$ GeV/c. 
%In 2008 the TPC read-out and Data Aqusition system were upgraded to increase the event recording rate by a factor of $\approx$ 10 (important i.e. for high $p_T$ particles). 
In 2011 the NA49 Forward Calorimeter was replaced by the Projectile Spectator Detector (PSD) with a five times better resolution ($\sigma(E)/E \approx 0.55/ \sqrt{E/(1GeV)}$), which corresponds to one nucleon(!) resolution in the studied energy range. Such precise measurement of the energy of projectile spectators is needed for a tight centrality selection as required for the analysis of fluctuations. The PSD will also allow to reconstruct the event reaction plane. For the 2011 $Be$ runs 32 of 44 PSD modules were ready, sufficient to cover projectile spectators at the three highest SPS energies. The remaining modules will be finished in 2012 for $Be$ runs at lower energies.    
Channeling of a high intensity heavy ion beam through the gas volume of the Vertex TPCs has limitations when compared to a proton beam. Delta electrons produced in the gas volume inside the VTPCs from heavy ion beam-gas interactions (electrons kicked off from TPC gas atoms by beam ions/spectators) may significantly increase the background in the TPCs and distort measurements of event-by-event fluctuations. Therefore in 2011 helium filled beam pipes were installed inside both VTPCs (around the beam line) resulting in a reduction of the $\delta$-electron background by a factor of 10. 

%In NA61 $Be+Be$, $Ar+Ca$, and $Xe+La$ collisions will be registered. The beams of $Ar$ and $Xe$ ions will be delivered directly by SPS, whereas production of fragmentation beam is neceserly for $Be+Be$ interactions. Secondary ion beam in NA61 is produduced by colliding primary $Pb$ beam with fragmentation target. Its length is optimized to the desired fragment production (for example $Be$). The produced $Pb$ beam fragments are then analyzed by use of the Ion Fragment Separator which allows to select clean ion beams. Fragment separator is a double magnetic spectrometer to separate ion fragments corresponding to selected magnetic rigidity $B\rho$. Additionally it includes the degrader ($Cu$ plate where ions lose energy $dE/dx \sim Z^2$) which allow to reach required beam purity. The procedure was sucessfully tested in 2010 for 13.9$A$ and 80$A$ GeV $Pb$ ion beams and used in 2011 for $Pb$ ions at 40$A$, 80$A$, and 158$A$ GeV ($ ^{7}Be$ beams were produced).  
    
%Ready for 2011 Be run: Z-detectors (measure ion charge for on-line selection of secondary ions, A-detector (measures mass composition of secondary ion beam). Low Momentum Particle Detector (LMPD), for centrality determination in p+A, is ready and the pilot p+Pb data were registered in 2011   

\section{Results}

\subsection{Particle spectra in p+C at 31 GeV}

\begin{figure}
\includegraphics[width=0.32\textwidth]{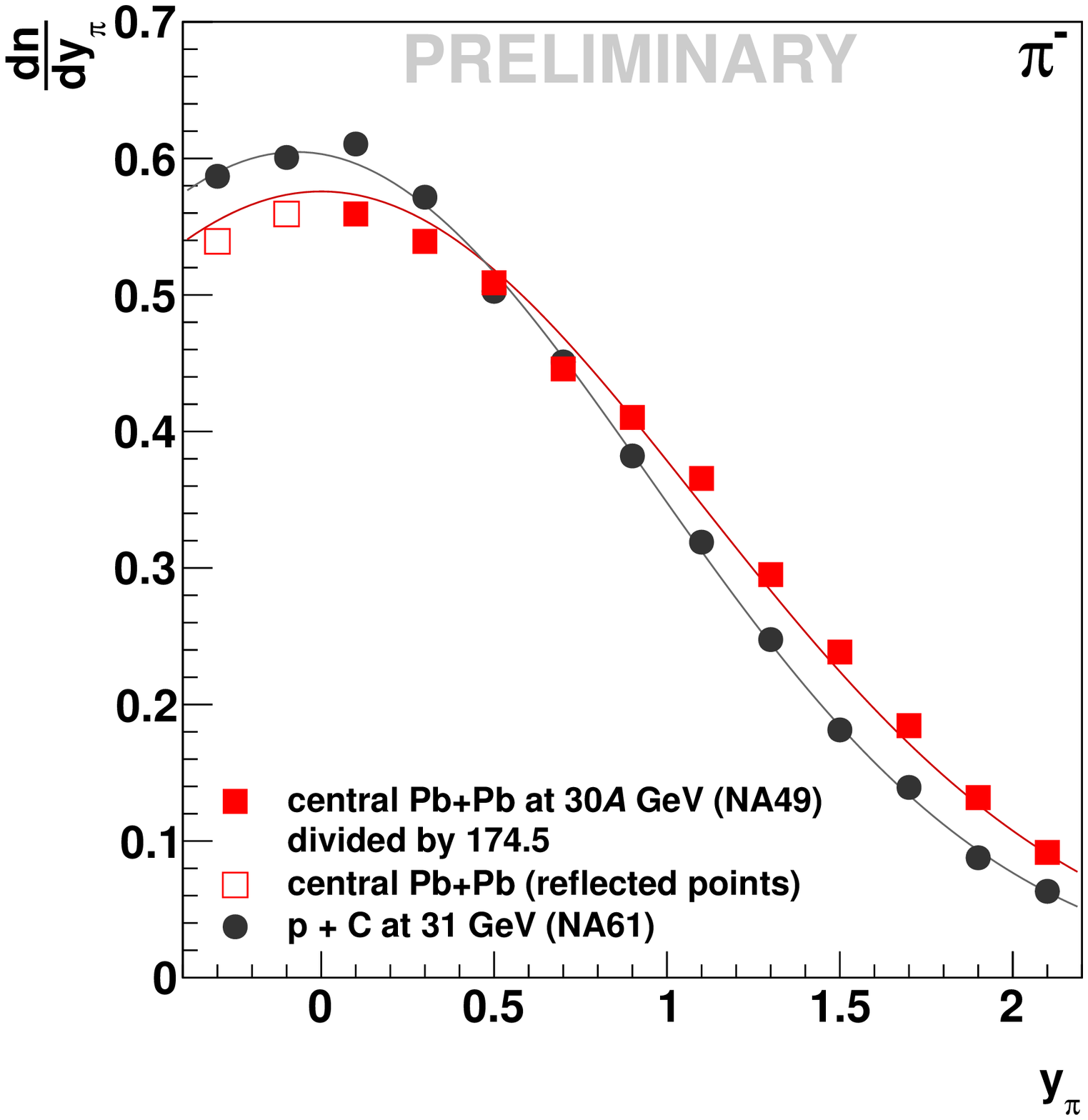}
\includegraphics[width=0.32\textwidth]{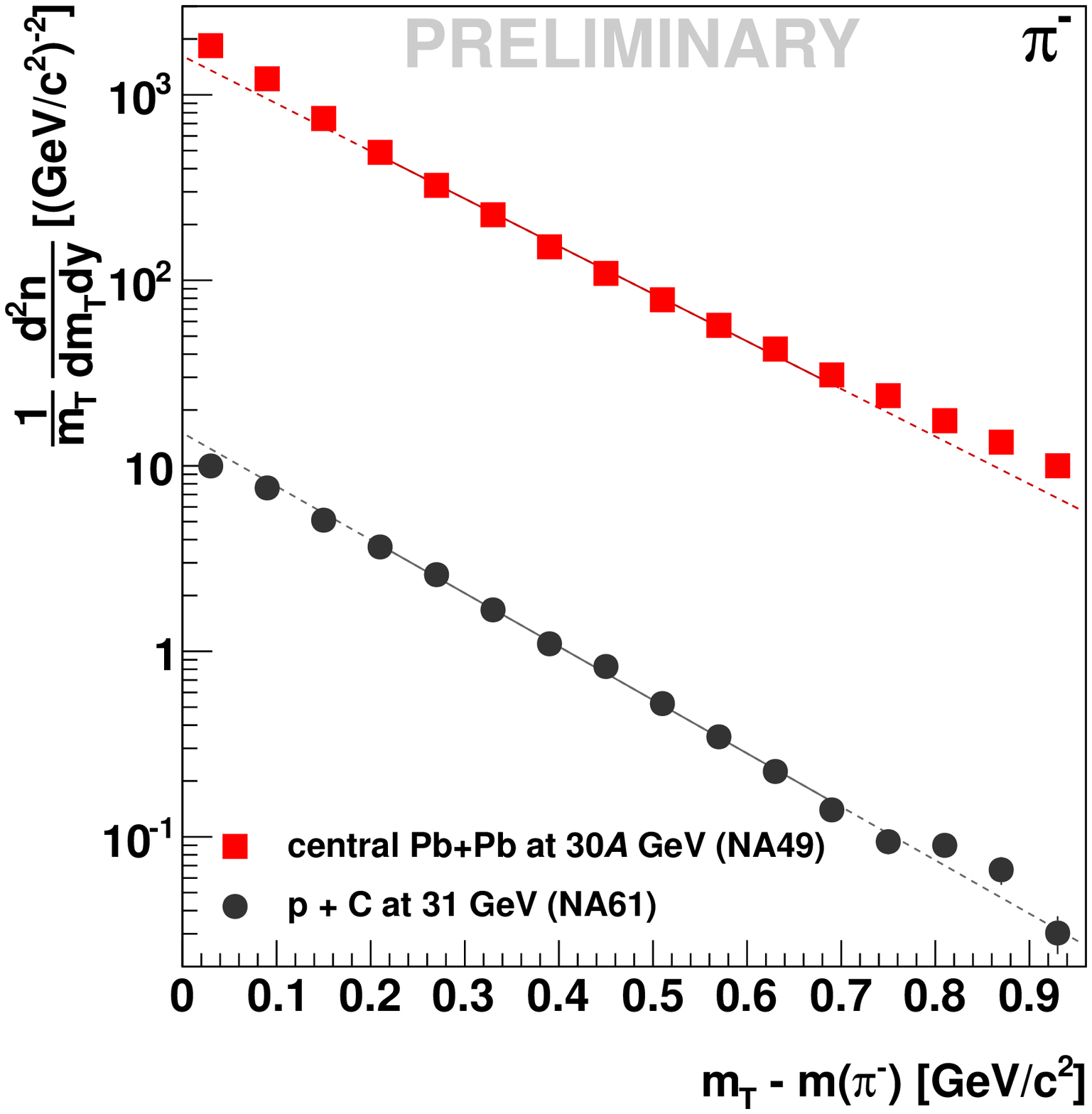}
\includegraphics[width=0.32\textwidth]{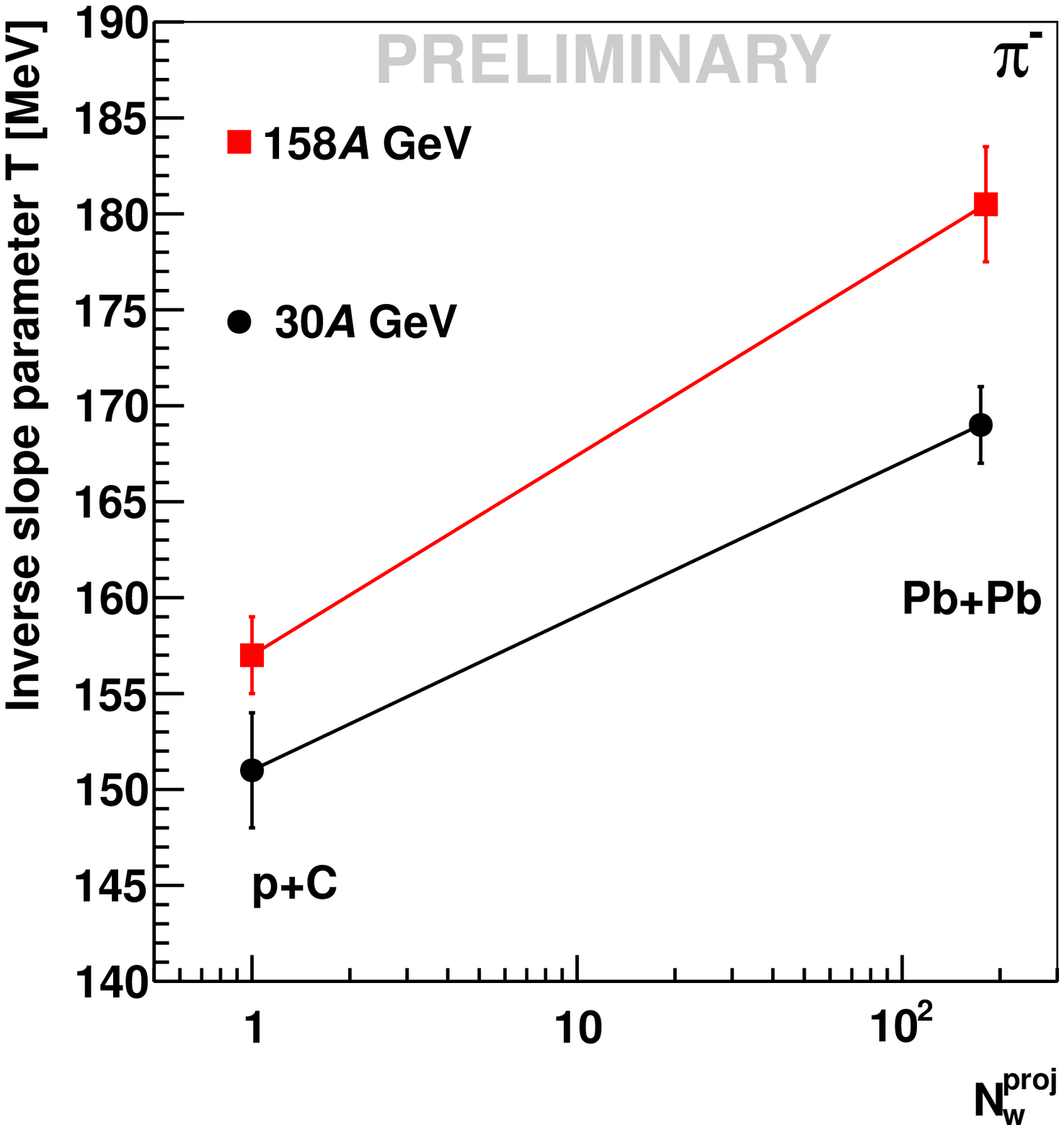}
\vspace{-0.5cm}
\caption{Rapidity distributions of $\pi^{-}$ in the center-of-mass reference system (left). Transverse mass spectra of $\pi^{-}$ (middle). Inverse slope parameters of negative pions $m_T$ spectra versus $N_{W}^{proj}$ (right).}
\label{pion_spectra}
\end{figure}

First $\pi^{-}$ results relevant to the study of the properties of the onset of deconfinement are presented in Fig. \ref{pion_spectra}. The left panel shows the rapidity spectrum from p+C interactions at 31 GeV compared to NA49 results from central Pb+Pb collisions at 30$A$ GeV \cite{na49_kpi}. Pb+Pb points are scaled by the mean number of wounded projectile nucleons. The $p+C$ rapidity spectrum is shifted towards target rapidity with respect to $Pb+Pb$ reactions due to the projectile-target asymmetry of the initial state. The NA61 $p+C$ results at the energy of the OD confirm the approximate proportionality of the pion yield to the mean number of wounded nucleons. 

The middle panel of Fig. \ref{pion_spectra} shows mid-rapidity ($0<y_{\pi}<0.2$) transverse mass spectra of pions in p+C interactions at 31 GeV and in central Pb+Pb collisions at 30$A$ GeV \cite{na49_kpi}. The shape of $m_T$ spectra changes from a convex form in $p+C$ to a concave one in central $Pb+Pb$ (with respect to corresponding exponential fits). In a hydrodynamical approach this is due to strong radial collective flow in $Pb+Pb$ collisions, which is absent in $p+C$ interactions. An exponential function was fitted to the transverse mass spectra in the range $0.2 < m_T -m_{\pi}<0.7$ GeV/c$^2$ (the same range was used by NA49 \cite{na49_kpi}). The fitted inverse slope parameter $T$ for $p+C$ interactions at 31 GeV is plotted in Fig. \ref{pion_spectra} (right) together with the slopes for $p+C$ interactions at 158 GeV \cite{pC158} and for central $Pb+Pb$ at 30$A$ and 158$A$ GeV \cite{na49_kpi}. As seen the inverse slope parameter increases with the collision energy, and with the number of wounded projectile nucleons. In hydrodynamical models this can be interpreted as an increase of radial flow and/or temperature.

The $p+C$ data at 31 GeV are also analyzed to obtain $\Lambda$, $K^0_s$ and $\Delta^{++}$ spectra. The secondary vertex (V0) technique is used to analyze $\Lambda$ and $K^0_s$, whereas $\Delta^{++}$ is searched by combining all possible ($p,\pi^{+}$) pairs originating from the main $p+C$ interaction vertex (the combinatorial background is obtained by the event mixing method). Identification of $p$ and $\pi^{+}$ uses measurements of $dE/dx$ in the TPCs and time of flight in the ToF detectors. First results for $\Lambda$, $K^0_s$ and $\Delta^{++}$ were already obtained.
%During the conference examples of fitted $\Lambda$, $K^0_s$ and $\Delta^{++}$ invariant mass distributions were shown.     

\subsection{Event-by-event fluctuations in p+C at 31 GeV}

\begin{figure}
\vspace{-0.5cm}
\includegraphics[width=0.32\textwidth]{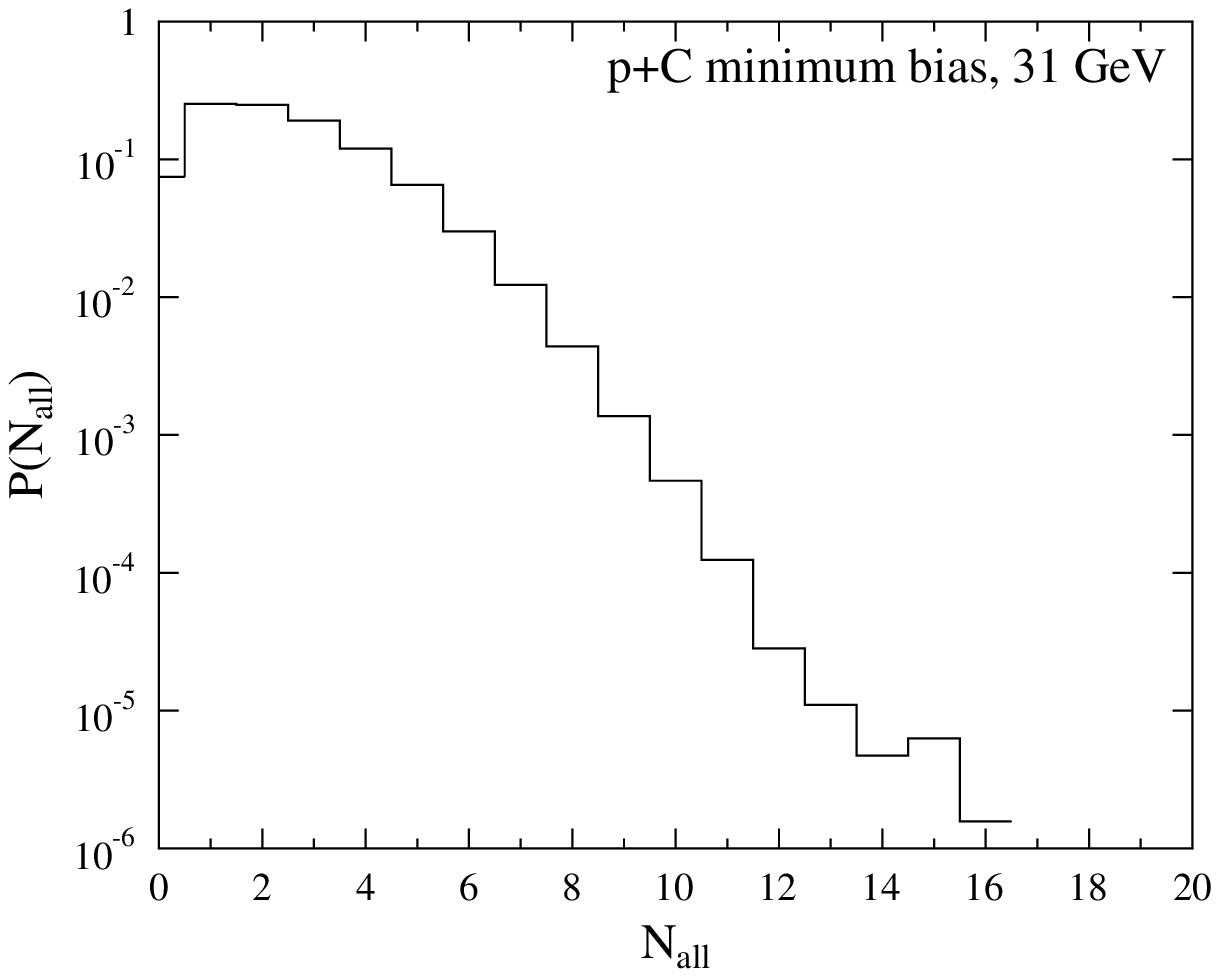}
\includegraphics[width=0.28\textwidth]{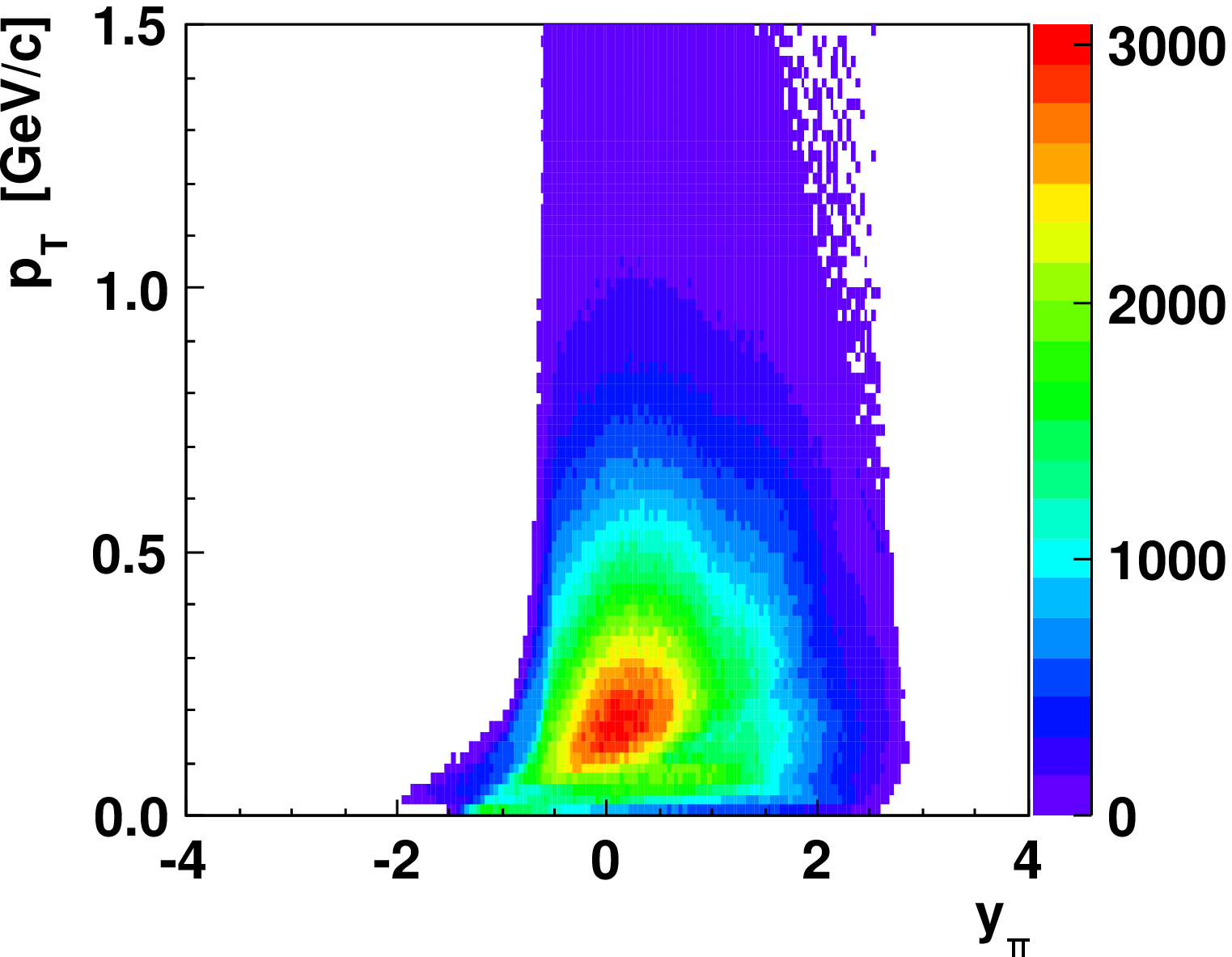}
\includegraphics[width=0.28\textwidth]{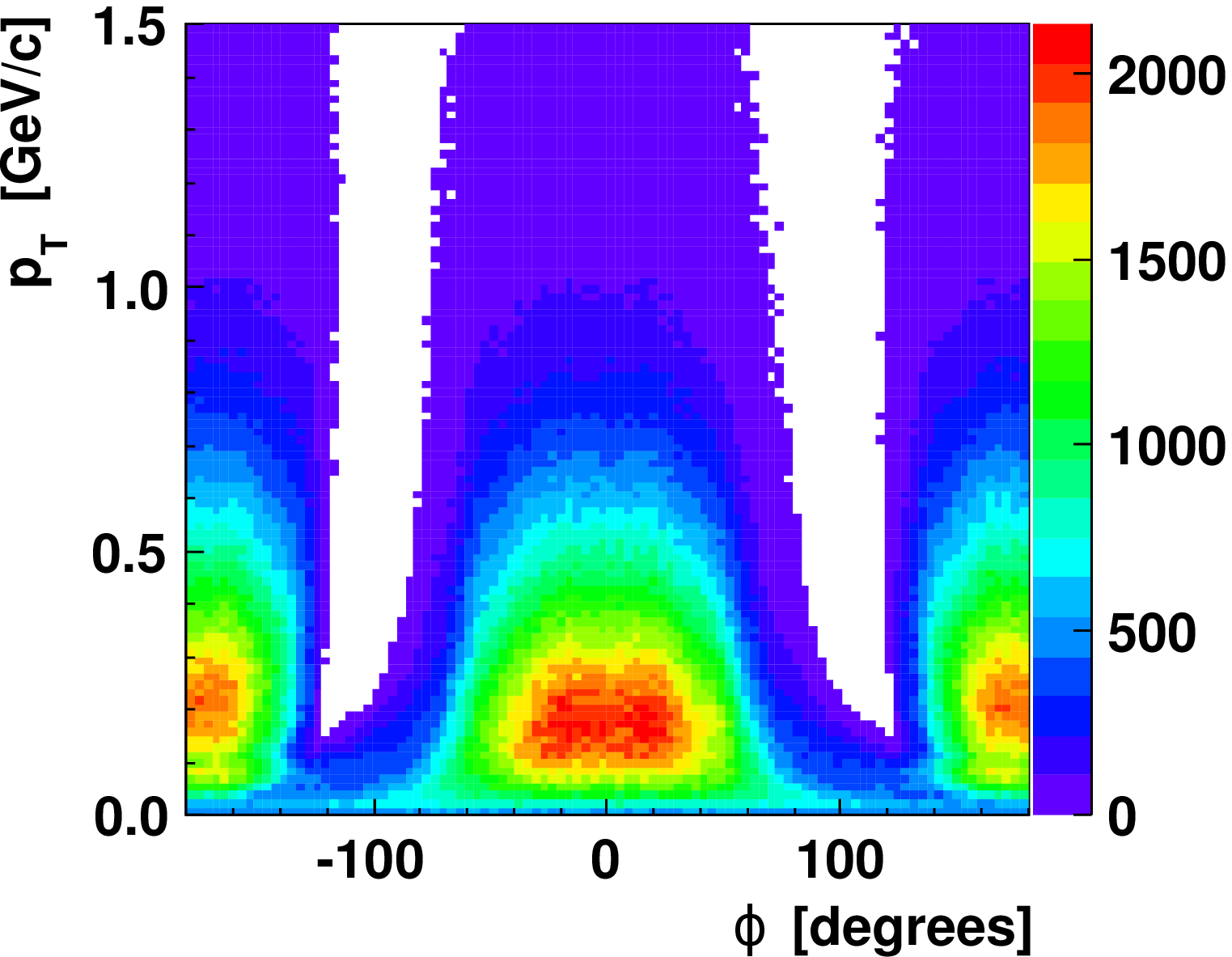}
\vspace{-0.5cm}
\caption{Multiplicity distribution of charged particles in $p+C$ at 31 GeV (left). Acceptance used in e-by-e analysis: transverse momentum versus rapidity of charged particles (with pion mass) and versus azimuthal angle (middle, right).}
\label{acceptance}
\end{figure}

First results on e-by-e multiplicity and average $p_T$ fluctuations were obtained from studying the scaled variance of the multiplicity distribution $\omega$ and the $\Phi_{p_T}$ measure (see \cite{kg_qm09} for details). The results are corrected for the effect of contamination from non-target interactions (see \cite{cpod10_TCKG} for complete procedure), however they are not yet corrected for on-line and off-line event selection biases. The results were obtained in the acceptance \footnote {Please note, that the measurement of average $p_T$ and multiplicity fluctuations in NA49 was limited to the very forward rapidity ($y_{\pi}>1$) \cite{kg_qm09}. Here, we successfully used complete rapidity range as available in NA61.} as shown in Fig. \ref{acceptance}.

Multiplicity distributions (i.e. Fig. \ref{acceptance} (left)) for all three charge combinations are close to Poisson. The scaled variance $\omega=\frac {\langle N^2 \rangle - \langle N \rangle ^2 }{ \langle N \rangle }$ equals $1.08 \pm 0.03$ for all charged, $0.93 \pm 0.03 $ for negatively charged and $0.83 \pm 0.03$ for positively charged particles (statistical errors only). The mean multiplicities in the studied kinematic range are approximately 2.45, 0.87, and 1.58 for all charged, negatively charged, and positively charged particles, respectively. 
The $\Phi_{p_T}$ measure in $p+C$ interactions at 31 GeV was found to be $9.35 \pm 0.15$ MeV/c for all charged, $1.60 \pm 0.15$ MeV/c for negatively charged, and $1.81 \pm 0.16$ MeV/c for positively charged particles. The estimated systematic errors are not higher than 0.6-1.0 MeV/c (depending on charge selection). The stronger correlation for all charged particles is reproduced by UrQMD2.3 for which values of \footnote{In the UrQMD model no NA61 acceptance filter was applied (yet)} $10.8 \pm 0.9$ MeV/c, $-0.7 \pm 0.8$ MeV/c and $4.3 \pm 0.9$ MeV/c are obtained for all charged, negatively charged and positively charged, respectively.

\section{Summary}

The NA61 experiment performs measurements for three different topics: critical point and onset of deconfinement (ion program), neutrino and cosmic-ray physics, and high $p_T$ particle production. The ion program aims to discover the critical point of strongly interacting matter and guarantees systematic data on the onset of deconfinement. The 2D scan of the phase diagram is in progress and will completed by the end of 2015. During the conference first results on $p+C$ interactions at 31 GeV were presented. The analysis of the $p+p$ energy scan is in progress.

\end{document}